\documentstyle[aps,epsf]{revtex}

\newcommand{\eq}{\begin{equation}}
\newcommand{\en}{\end{equation}}
\newcommand{\eqn}{\begin{eqnarray}}
\newcommand{\enn}{\end{eqnarray}}
\newcommand{\nn}{\nonumber }
\newcommand{\beq}{\begin{equation}}
\newcommand{\eeq}{\end{equation}}

\begin{document}
\title{\bf KINEMATICS OF $AdS_{5}/CFT_{4}$
DUALITY }
\author{D. Minic\footnote{Talk given at DPF '99; UCLA, Jan. 7, 1999. This research was partially supported
by the US. Department of Energy under grant number DE-FG03-84ER40168.
e-mail: minic@physics.usc.edu }}
\address{Department of
Physics and Astronomy, 
University of Southern California,
Los Angeles, CA 90089-0484}
\maketitle

\begin{abstract}
We review the construction of the unitary supermultiplets of the ${\cal{N}}=8$ 
$d=5$ 
anti-de Sitter ($AdS_5$)
superalgebra $SU(2,2|4)$, which is the symmetry group of type IIB superstring 
theory
on $AdS_5 \times S^5$, using the oscillator method. \ 
\end{abstract}


\baselineskip 14pt

%


\section{Introduction}


Recently a great deal of work has been done on AdS/CFT
(anti-de Sitter/conformal field theory) dualities in various
dimensions. This activity 
was primarily started with
the original conjecture of Maldacena \cite{mald}, \cite{pol,witt} about the
relation between the large $N$ limits of certain conformal field theories
in $d$ dimensions to M-theory/string theory
compactified to $d+1$-dimensional AdS spacetimes. 
The prime example of 
AdS/CFT duality is the duality between the large $N$ limit
of ${\cal{N}}=4$ $SU(N)$ super Yang-Mills theory in $d=4$ and type IIB 
superstring theory on $AdS_5 \times S^5$.

In this lecture we will review some results obtained in
collaboration with M. G\"{u}naydin and M. Zagermann \cite{gmz1}, \cite{gmz2} regarding the construction of the unitary supermultiplets of the ${\cal{N}}=8$ $d=5$ 
anti-de Sitter ($AdS_5$)
superalgebra $SU(2,2|4)$ ,
which is the symmetry group of type IIB superstring theory on $AdS_5 \times S^5$, using the oscillator method.
Our results should be relevant for the understanding of the spectrum of
type IIB string theory on $AdS_5 \times S^{5}$ \cite{kallosh}.

\section{Short Review of the Oscillator Method}


In \cite{mgcs} a general oscillator method was developed for
constructing the unitary irreducible representations (UIR) of
the lowest (or highest) weight type of non-compact groups.
The oscillator method yields the UIR's of lowest weight type of a 
noncompact group over the Fock space of a set of bosonic 
oscillators. To achieve this one realizes the generators of the 
noncompact group as bilinears of  sets of bosonic oscillators
transforming  in a finite dimensional representation of its maximal
compact subgroup. The minimal realization of these generators requires
either one or two sets of bosonic annihilation and 
creation operators transforming
 irreducibly under its maximal compact subgroup. These minimal 
 representations are 
fundamental in that all the other ones can be obtained from the minimal
representations by a simple tensoring procedure.  
 
These fundamental representations
are nothing but a generalization of the celebrated remarkable representations
of the $AdS_4$ group $SO(3,2)$ discovered by Dirac
\cite{pam}
long time ago, which
were later named singletons \cite{fron} (indicating
the fact that the remarkable representations of Dirac 
corresponding to the fields living on the boundary of
$AdS_4$ are singular when the Poincare limit is taken).
In the language of the oscillator method, these singleton representations
require a single set of oscillators transforming in the fundamental
representation of the maximal compact subgroup of the covering group
$Sp(4,R)$ of $SO(3,2)$  \cite{mg81}, \cite{mgnw} (a fact that meshes nicely
with the name singleton).
In some cases (as with the $AdS_5$ group
$SU(2,2)$) the fundamental representations require two sets of oscillators,
and they were called doubletons in  \cite{mgnm}, \cite{gnw}. The general oscillator 
construction of the lowest (or highest) weight representations 
of non-compact supergroups (i.e. the case when the even subgroup
is non-compact) was given in \cite{ibmg}. 
 The oscillator method was further developed and applied to the spectra
of Kaluza-Klein supergravity theories in references \cite{mgnm}, 
\cite{mgnw}, \cite{gnw}.

A non-compact group $G$ that admits unitary representations of the
lowest weight type has a maximal compact subgroup $G^{0}$  of the form
$G^{0} = H \times U(1)$ with respect to whose Lie algebra $g^0$  one has
a three grading of the Lie algebra $g$ of $G$,
\eq
g = g^{-1} \oplus g^{0} \oplus g^{+1}
\en
which simply means that the commutators of elements of grade
$k$ and $l$ satisfy
\eq
[g^{k},g^{l}] \subseteq g^{k+l}.
\en
Here $g^{k+l} = 0$ for $|k+l| >1$.

For example, for $SU(1,1)$ this corresponds to the standard
decomposition $g = L_{+} \oplus L_{0} \oplus L_{-}$ where
\eq
[L_{0},L_{\pm}] = \pm L_{\pm}, \quad [L_{+},L_{-}] =2L_{0}.
\en
The three grading is determined by the generator $E$ of the
$U(1)$ factor of the maximal compact subgroup
\eq
[E,g^{+1}] =g^{+1}, \quad [E,g^{-1}] =-g^{-1}, \quad
[E,g^{0}] = 0.
\en
In most physical applications $E$  turns out
to be the energy operator.
In such cases the unitary lowest weight representations
correspond to positive energy representations.

The  bosonic annihilation
and creation operators in terms of which one realizes the generators
of $G$ transform typically in the fundamental
and its conjugate representation of $H$.
In the Fock space $\cal{H}$ of all the oscillators one
chooses a set of  states
$|\Omega \rangle$ which transform irreducibly under $H \times U(1)$
and are annihilated by all the generators in $g^{-1}$.
Then by acting on $|\Omega \rangle$ with generators in
$g^{+1}$ one obtains an infinite set of states
\eq
|\Omega \rangle ,\quad  g^{+1}|\Omega \rangle ,\quad
g^{+1} g^{+1}|\Omega \rangle , ...
\en
which form an UIR of the lowest weight (positive energy) type
of $G$. The infinite set of states thus obtained corresponds to the
decomposition of the UIR of $G$ with respect to its maximal
compact subgroup.

As we have already emphasized, whenever
 we can realize the generators of $G$ in
terms of a single set of bosonic creation (and annihilation )
operators transforming in an irreducible representation (and its conjugate) of
the compact subgroup $H$ then the corresponding
UIRs  will be called singleton representations
and  there exist two such representations for a given group
 $G$. For the AdS group in $d=4$  the singleton
representations correspond to scalar and spinor fields .
In certain cases  we need  a minimum of two sets of 
bosonic creation  and annihilation 
operators transforming irreducibly under $H$ to realize the
generators of $G$. In such cases  the
corresponding UIRs  are called doubleton
representations and there exist
infinitely many doubleton representations
of $G$ corresponding to fields of different
"spins".

Even though the Poincare limit of the singleton (or doubleton) representations
is singular, the tensor product of two singleton (or doubleton)
representations decomposes into an infinite set of "massless"
irreducible representations which do have a smooth Poincare limit
\cite{fron}, \cite{mg81}, \cite{mgnw}. 
Furthermore, tensoring more than two singletons or doubletons
representations leads to "massive" representations of
AdS groups and supergroups.

The relation between Maldacena's
conjecture and the dynamics of the singleton and doubleton fields that
live on the boundary of AdS 
spacetimes was reviewed in \cite{sfcf}, \cite{mgdm}, \cite{dupo}.

\section{The superalgebra $SU(2,2|4)$ }

The centrally extended symmetry supergroup of type IIB superstring theory on $AdS_{5}\times S^{5}$
is the supergroup $SU(2,2|4)$ with the even subgroup
$SU(2,2) \times SU(4) \times U(1)_Z$, where $SU(4)$ is the double cover of 
$SO(6)$, the isometry group of the five sphere \cite{mgnm}. The Abelian $U(1)_Z$
generator, which we will call $Z$,
commutes with all the other
generators and acts like a central charge. Therefore, $SU(2,2|4)$
is not a simple Lie superalgebra. By factoring out this Abelian
ideal one obtains a simple Lie superalgebra, denoted by $PSU(2,2|4)$,
whose even subalgebra is simply $SU(2,2)\times SU(4)$.
We consider below the centrally extended supergroup $SU(2,2|4)$
\cite{gmz1}. The representations 
of $PSU(2,2|4)$ correspond simply to representations of $SU(2,2|4)$
with $Z=0$. We should also note that
both $SU(2,2|4)$ and $PSU(2,2|4)$ admit an outer automorphism $U(1)_Y$
that can be identified with the $U(1)$ subgroup of the $SU(1,1)$ 
symmetry of IIB supergravity in $d=10$ \cite{mgnm}.
$SU(2,2|4)$ can be interpreted as the ${\cal{N}}=8$ extended AdS superalgebra
in $d=5$ or as the ${\cal{N}}=4$ extended conformal superalgebra in $d=4$.

The algebra of ${\cal{N}}$-extended conformal supersymmetry  in $d=4$  can 
be written in a covariant  form as follows 
($i,j =1, \ldots ,{\cal{N}}$; $a,b =0,1,2,3,5,6$)\cite{sohn}:
\begin{eqnarray}
[\Xi_{i}, M_{ab}] =  \Sigma(M_{ab}) \Xi_{i},& &
[{\bar{\Xi}}^{i}, M_{ab}] = - {\bar{\Xi}}^{i}\Sigma(M_{ab})\nonumber\\
\{ \Xi_{i}, \Xi_{j} \} = 
\{ {\bar{\Xi}}^{i}, {\bar{\Xi}}^{j} \} =0, & &
\{ \Xi_{i}, {\bar{\Xi}}^{j} \} = 
2{\delta}^{j}_{i} \Sigma(M^{ab}) M_{ab} -4B^j_i\nonumber\\
{[}B^j_i, M_{ab}{]}=0, & & {[}B^j_i, B^l_k{]}={\delta}^l_i B^j_k - 
{\delta}^j_k B^l_i\nonumber\\
{[}\Xi_{i}, B^k_j{]} = {\delta}^k_i \Xi_{j}  - {1 \over 4} 
{\delta}^k_j \Xi_{i}, & & 
{[}{\bar{\Xi}}^{i}, B^k_j{]} = -{\delta}^i_j {\bar{\Xi}}^{k}  + {1 \over 4} 
{\delta}^k_j {\bar{\Xi}}^{i},
\end{eqnarray}
where the (four component) conformal spinor $\Xi$ is defined in terms of the
the chiral components of the Lorentz spinors $Q$ and $S$ (the generators of
Poincar\'{e} and $S$ type supersymmetry) as
\eq
\Xi \equiv \left(\matrix{Q_{\alpha} \cr
                        {\bar{S}}^{\dot{\alpha}} \cr} \right).
\en
The $B_i^j$ are the generators of the internal (R-)symmetry group 
$U({\cal{N}})$ and the  $\Sigma(M_{ab})$ are $4 \times 4$ matrices generating an irreducible representation of $SU(2,2)$ \cite{gmz1}, \cite{sohn}.

The superalgebra $SU(2,2|4)$ has a three graded decomposition with
respect to its compact subsuperalgebra $SU(2|2)\times SU(2|2) \times U(1)$
\eq
g = L^{+} \oplus L^{0} \oplus L^{-},
\en
where
\eq
[L^{0},L^{\pm}] \subseteq  L^{\pm}, \quad
[L^{+},L^{-}] \subseteq L^{0}, \quad
[L^{+},L^{+}] = 0=[L^{-},L^{-}].
\en
Here $L^{0}$ represents the generators of
$SU(2|2) \times SU(2|2) \times U(1)$.

The Lie superalgebra $SU(2,2|4)$
can be realized in terms of bilinear combinations of bosonic and
fermionic annihilation and creation operators $\xi_{A}$
($\xi^{A}={\xi_{A}}^{\dagger}$) and $\eta_{M}$
($\eta^{M}={\eta_{M}}^{\dagger}$)
which transform covariantly and contravariantly
under the  two $SU(2|2)$ subsupergroups of $SU(2,2|4)$ 
\cite{gmz1,mg81,ibmg,mgnm}
\eq
\xi_{A} = \left(\matrix{a_{i} \cr
                        \alpha_{\gamma} \cr} \right) ,\quad
\xi^{A} = \left(\matrix{a^{i} \cr
                        \alpha^{\gamma} \cr} \right)
\en
and
\eq
\eta_{M} = \left(\matrix{b_{r} \cr
                        \beta_{x} \cr} \right) , \quad
\eta^{M} = \left(\matrix{b^{r} \cr
                        \beta^{x} \cr} \right)
\en
with $i,j=1,2$; $\gamma,\delta=1,2$; $r,s=1,2$; $x,y=1,2$ and
\eq
[a_i, a^j] = \delta_{i}^{j} , \quad
\{\alpha_{\gamma}, \alpha^{\delta}\} = \delta_{\gamma}^{\delta}
\en
\eq
[b_r, b^s] = \delta_{r}^{s} , \quad
\{\beta_{x}, \beta^{y}\} = \delta_{x}^{y}.
\en
Annihilation and creation operators are labelled by lower and
upper indices, respectively. 
The generators of $SU(2,2|4)$ are given in terms of the above 
superoscillators schematically as
\eq
L^{-} = {\vec{\xi}}_{A} \cdot {\vec{\eta}}_{M}, \quad
L^{0} = {\vec{\xi}}^{A} \cdot {\vec{\xi}}_{B}
\oplus {\vec{\eta}}^{M} \cdot {\vec{\eta}}_{N}, \quad
L^{+} ={\vec{\xi}}^{A} \cdot {\vec{\eta}}^{M},\label{xieta}
\en
where the arrows over $\xi$ and $\eta$ again indicate that we are taking an
arbitrary number $P$ of ``generations'' of
superoscillators and the dot represents the summation
over the internal index $K = 1,...,P$, i.e
${\vec{\xi}}_{A} \cdot {\vec{\eta}}_{M} \equiv \sum_{K=1}^{P}
{\xi_{A}}(K){\eta_{M}}(K)$.

The even subgroup $SU(2,2)\times SU(4)\times U(1)_Z$ is obviously generated 
by the di-bosonic and di-fermionic generators. In particular, one
recovers the  $SU(2,2)$ generators in terms of the bosonic 
oscillators:
\eqn
L^{j}_{i} &=& {\vec{a}}^{j} \cdot {\vec{a}}_{i}
-{1 \over 2} \delta^{j}_{i}N_{a}, \quad
R^{r}_{s} = \vec{b^{r}} \cdot \vec{b_{s}}
-{1 \over 2} \delta^{r}_{s} N_{b}\cr
E &=& \frac{1}{2} \{ {\vec{a}}^{i} \cdot {\vec{a}}_{i} + {\vec{b}}_{r} 
\cdot {\vec{b}}^{r}\}=
{1 \over 2} \{ N_a +  N_b +2P  \}\cr
L^{-}_{ir} &=&{\vec{a}}_{i} \cdot {\vec{b}}_{r}, \quad
L^{+ ri}={\vec{b}}^{r} \cdot {\vec{a}}^{i}\label{su22}
\enn
satisfying
\eq
[L^{-}_{ir},L^{+ sj}]
= \delta^{s}_{r} L^{j}_{i} + \delta^{j}_{i} R^{s}_{r}
+ \delta^{j}_{i}\delta^{s}_{r} E. 
\en
Here, $N_{a} \equiv {\vec{a}}^{i} \cdot {\vec{a}}_{i}, 
N_{b} \equiv {\vec{b}}^{r} \cdot {\vec{b}}_{r}$ are the
bosonic number operators.

Similarly, the $SU(4)$ generators in their $SU(2)\times SU(2) \times U(1)$ 
basis are expressed  in 
terms of the fermionic oscillators $\alpha$ and $\beta$:
\eqn
A^{\delta}_{\gamma} &=& {\vec{\alpha}}^{\delta} \cdot {\vec{\alpha}}_{\gamma}
-{1 \over 2} \delta^{\delta}_{\gamma} N_{\alpha}, \quad
B^{y}_{x} = {\vec{\beta}}^{y} \cdot {\vec{\beta}}_{x}
-{1 \over 2} \delta^{y}_{x} N_{\beta} \cr
C &=& \frac{1}{2}\{-{\vec{\alpha}}^{\delta} \cdot {\vec{\alpha}}_{\delta}
+ \vec{\beta_{x}} \cdot {\vec{\beta}}^{x}\}={1 \over 2} \{- N_{\alpha} -  
N_{\beta} +2P\} \cr
L^{-}_{\gamma x} & = &{\vec{\alpha}}_{\gamma} \cdot {\vec{\beta}}_{x},
\quad L^{+ x\gamma}={\vec{\beta}}^{x} \cdot 
{\vec{\alpha}}^{\gamma}\label{su4}
\enn
with the closure relation
\eq
[L^{-}_{\gamma x},L^{+ y \delta}]
= -\delta^{y}_{x} A^{\delta}_{\gamma} - \delta^{\delta}_{\gamma} B^{y}_{x}
+ \delta^{y}_{x} \delta^{\delta}_{\gamma} C.
\en
Here $N_{\alpha}={\vec{\alpha}}^{\delta} \cdot {\vec{\alpha}}_{\delta} $ 
and $N_{\beta} = \vec{\beta^{x}} \cdot {\vec{\beta}}_{x}$
are the fermionic number operators.

Finally, the central charge-like $U(1)_Z$ generator $Z$ is given by
\eq 
Z=\frac{1}{2}\{ N_{a}+N_{\alpha}-N_{b}-N_{\beta} \}.
\en
Analogously, the odd generators are given by products of bosonic and 
fermionic oscillators and satisfy the following closure relations  
\eqn
\{ {\vec{a}}_{i} \cdot {\vec{\beta}}_{x},{\vec{\beta}}^{y}
\cdot {\vec{a}}^{j} \}
&=& \delta^{y}_{x} L^{j}_{i} - \delta^{j}_{i} B^{y}_{x}
+ {1 \over 2}\delta^{y}_{x}\delta^{j}_{i} {(E+C+Z)} \cr
\{ {\vec{\alpha}}_{\gamma} \cdot {\vec{b}}_{r},
{\vec{b}}^{s} \cdot {\vec{\alpha}}^{\delta} \}
&=& -\delta^{s}_{r} A^{\delta}_{\gamma} + \delta^{\delta}_{\gamma} R^{s}_{r}
+ {1 \over 2}\delta^{s}_{r} \delta^{\delta}_{\gamma} {(E+C-Z)} \cr
\{ {\vec{a}}^{i} \cdot {\vec{\alpha}}_{\gamma},{\vec{\alpha}}^{\delta}
\cdot {\vec{a}}_{j}\}
&=& \delta^{\delta}_{\gamma} L^{i}_{j} + \delta^{i}_{j} A^{\delta}_{\gamma}
+ {1 \over 2}\delta^{\delta}_{\gamma} \delta^{i}_{j} {(E-C+Z)} \cr
\{ {\vec{b}}^{r} \cdot {\vec{\beta}}_{x},{\vec{\beta}}^{y}
\cdot {\vec{b}}_{s} \}
&=& \delta^{y}_{x} R^{r}_{s} + \delta^{r}_{s} B^{y}_{x}
+ {1 \over 2}\delta^{y}_{x}\delta^{r}_{s} {(E-C-Z)}.
\enn
 
The generator $Y$ of the outer automorphism group $U(1)_Y$ is simply
\eq
Y= N_{\alpha}  - N_{\beta}.
\en

\section{Unitary Supermultiplets of $SU(2,2|4)$}

To construct a basis for a lowest weight UIR of $SU(2,2|4)$,
one starts from a set of states, collectively denoted by $ |\Omega \rangle$, 
in the  Fock space of the oscillators $a$, $b$, $\alpha$, $\beta$ that
transforms irreducibly under $SU(2|2) \times
SU(2|2) \times U(1)$ and that is annihilated by all the generators 
${\vec{\xi}}_{A}\cdot {\vec{\eta}}_{M}\equiv ({\vec{a}}_{i} \cdot {\vec{b}}_{r}
\oplus {\vec{a}}_{i}~\cdot~{\vec{\beta}}_{x}
\oplus {\vec{\alpha}}_{\gamma}~\cdot~{\vec{b}}_{r}
\oplus {\vec{\alpha}}_{\gamma} \cdot {\vec{\beta}}_{x})$ of $L^{-}$
\eq
{\vec{\xi}}_{A}\cdot {\vec{\eta}}_{M}|\Omega\rangle = 0.\label{lwv}
\en 
By acting
on  $ |\Omega \rangle$ repeatedly with $L^{+}$, 
one then
generates an infinite set of states that form a UIR of $SU(2,2|4)$

\eq
|\Omega \rangle ,\quad  L^{+1}|\Omega \rangle ,\quad
L^{+1} L^{+1}|\Omega \rangle , ...
\en
The irreducibility of the resulting representation of $SU(2,2|4)$ follows 
from the irreducibility of $|\Omega \rangle$ under $SU(2|2) \times
SU(2|2) \times U(1)$.
Because of the property (\ref{lwv}), $|\Omega\rangle$ 
as a whole
will be referred to as the ``lowest weight vector (lwv)'' of the
corresponding UIR of $SU(2,2|4)$.

In the restriction to the subspace involving purely bosonic
oscillators, the above construction reduces to the subalgebra $SU(2,2)$ 
and its positive energy UIR's. 
Similarly, when restricted to the subspace involving purely
fermionic oscillators, one gets the compact internal 
symmetry group $SU(4)$ (\ref{su4}),
and the above construction yields the representations of $SU(4)$ in
its $SU(2)\times SU(2) \times U(1)$ basis. 
Accordingly, a lowest weight UIR of $SU(2,2|4)$  decomposes into a 
direct sum of
finitely many positive energy UIR's of $SU(2,2)$ 
transforming in certain representations
of the internal symmetry group $SU(4)$. 
Thus each positive energy UIR of $SU(2,2|4)$
corresponds to a supermultiplet of fields living in $AdS_5$ or on its boundary.

\section{Doubleton Supermultiplets of $SU(2,2|4)$}

By choosing one pair of super oscillators ($\xi$ and $\eta$) in the
oscillator realization of $SU(2,2|4)$ (i.e. for $P=1$), one 
obtains the so-called 
doubleton supermultiplets. The doubleton supermultiplets contain only doubleton
representations of $SU(2,2)$, i.e. they are multiplets of fields living on 
the boundary of $AdS_{5}$ without a $5d$ Poincar\'{e} limit. Equivalently,
they can be characterized as multiplets of massless fields in $4d$ 
Minkowski space that form a UIR of the ${\cal{N}}=4$ superconformal algebra 
$SU(2,2|4)$.

The supermultiplet defined by the lwv $|\Omega \rangle = |0\rangle$ 
of $SU(2,2|4)$ 
is the unique irreducible CPT self-conjugate  doubleton supermultiplet. 
It is also the 
supermultiplet of ${\cal{N}}=4$ supersymmetric Yang-Mills theory in $d=4$ 
\cite{mgnm}.

If we take the following lowest weight vectors 
\eq
|\Omega \rangle = \xi^A |0\rangle \equiv a^i |0\rangle
\oplus \alpha^{\gamma} |0\rangle, \quad
|\Omega \rangle = \eta^A |0\rangle \equiv b^r |0\rangle
\oplus \beta^{x} |0\rangle
\en
we get a supermultiplet of spin range $3/2$ and its
CPT conjugate supermultiplet, respectively 
\cite{gmz1,gmz2}. 

These two doubleton supermultiplets of spin range 3/2 would occur in the 
${\cal{N}}=4$ super Yang-Mills theory if there is a well-defined  conformal 
(i.e. massless) limit of the 
$1/4$ BPS states described in ref \cite{ob}. 
These  $1/4$ BPS multiplets are massive representations of the four 
dimensional ${\cal{N}}=4$ Poincar\'{e} superalgebra
with two central charges, one of them saturating the BPS bound. As such, 
they are equivalent to  massive representations of the ${\cal{N}}=3$ 
Poincar\'{e} superalgebra without central charges. 
The corresponding multiplet with the lowest spin content (see e.g. \cite{jwjb})
contains 14 scalars, 14 spin 1/2 fermions, six vectors and one spin 3/2 
fermion,
giving altogether $2^{6}$ states. If a massless limit of such a multiplet 
exists, it should decompose into two self-conjugate doubleton multiplets
plus a doubleton supermultiplet of spin 3/2 plus its
CPT conjugate supermultiplet.

The lowest weight vectors of a generic doubleton supermultiplet of spin range 2
and its CPT-conjugate partner are \cite{gmz1,gmz2}
\eq
|\Omega \rangle = \xi^{A_1} \xi^{A_2}...\xi^{A_{2j}} |0\rangle, \quad
|\Omega \rangle = \eta^{A_1} \eta^{A_2}...\eta^{A_{2j}} |0\rangle.
\en

\section{``Massless'' supermultiplets of $SU(2,2|4)$}

The doubleton supermultiplets described in the last subsection are 
fundamental in the sense that all other lowest weight UIR's of $SU(2,2|4)$ 
occur in the 
tensor product of two or more doubleton supermultiplets. Instead of trying
to identify these irreducible submultiplets in the (in general reducible, but 
not 
fully reducible) tensor products, one simply increases the number $P$ of
oscillator generations so that the tensoring becomes implicit while the 
irreducibility stays manifest.

The simplest class, corresponding to $P=2$, contains the supermultiplets that 
are commonly referred to as ``massless'' in the $5d$ AdS sense. We will 
therefore use this as a name for all supermultiplets that are obtained by 
taking $P=2$ in the oscillator construction despite some problems with the 
notion of ``mass'' in AdS spacetimes \cite{gmz1}.

We will now give a complete list of the allowed $SU(2,2|4)$ lowest weight 
vectors $|\Omega \rangle$ for $P=2$.

The condition $L^{-}|\Omega \rangle = 0$ leaves the following possibilities:
\begin{itemize}
\item $|\Omega \rangle = |0\rangle$. This lwv gives rise to the $\mathcal{N}$ =
8 graviton supermultiplet in $AdS_{5}$ and occurs in the tensor product
of two CPT self-conjugate doubleton (i.e. $\mathcal{N}$ = 4 super Yang Mills) 
supermultiplets.
\item  $|\Omega \rangle = \xi^{A_1}(1) \xi^{A_2}(1) ... \xi^{A_{2j}}(1) 
|0\rangle $. The corresponding 
supermultiplets and also their conjugates resulting from 
\item $|\Omega \rangle = \eta^{A_1}(1) \eta^{A_2}(1) ... \eta^{A_{2j}}(1) 
|0\rangle $ have been listed in \cite{gmz1} (Tables 8 to 11). Increasing $j$ leads
to multiplets with higher and higher spins and AdS energies. For $j>3/2$ 
the spin range is always 4.
None of these multiplets can occur in the 
tensor product of two or more self-conjugate doubleton supermultiplets. They
require the chiral doubleton supermultiplets. 
\item $|\Omega \rangle = \xi^{A_1}(1) \xi^{A_2}(1) ... \xi^{A_{2j_{L}}}(1)
\eta^{B_1}(2) \eta^{B_2}(2) ... \eta^{B_{2j_{R}}} (2)|0\rangle$. The corresponding supermultiplets have been listed in \cite{gmz1} (Table 12). Again they involve spins and AdS energies that increase 
with $j_{L}$ and $j_{R}$, maintaining a constant spin range of 4  for 
$j_{L}, j_{R}\geq 1$.
\end{itemize}
In addition to these purely (super)symmetrized lwv's, one can also 
anti-(super)symmetrize pairs of superoscillators, since $P=2$. The requirement 
$L^{-}|\Omega\rangle =0$
then rules out the simultaneous appearance of $\xi$'s and $\eta$'s so that 
one is left with

\begin{itemize}
\item $|\Omega\rangle = 
\xi^{[A_{1}}(1)\xi^{B_{1}]}(2)...\xi^{[A_{n}}(1) \xi^{B_{n}]}(2) 
\xi^{C_{1}}(1)\dots \xi^{C_{k}}(1) |0\rangle$
\item $|\Omega\rangle = 
\eta^{[A_{1}}(1)\eta^{B_{1}]}(2)...\eta^{[A_{n}}(1) \eta^{B_{n}]}(2) 
\eta^{C_{1}}(1)\dots \eta^{C_{k}}(1) |0\rangle$ .
\end{itemize}
The special case $k=0$ then leads to the novel short multiplets listed in \cite{gmz2}. 
The simplest case is given by the following lowest weight
vectors
\eq
|\Omega \rangle =  
\xi^{[A_{1}}(1)\xi^{B_{1}]}(2)|0\rangle, \quad
|\Omega \rangle = \eta^{[A_{1}}(1)\eta^{B_{1}]}(2)|0\rangle
\en
describing a supermultiplet of spin range $2$ and its
CPT conjugate supermultiplet, respectively. 
Acting on $|\Omega\rangle$ with the supersymmetry generators 
${\vec{a}}^{i}\cdot 
{\vec{\beta}}^{x}$ and ${\vec{b}}^{r}\cdot{\vec{\alpha}}^{\gamma}$ of 
$L^{+}$ and collecting resulting $SU(2,2)\times SU(4)$ lwv's 
(i.e. states that are 
annihilated by ${\vec{a}}_{i}\cdot 
{\vec{b}}_{r}$ and ${\vec{\alpha}}_{\gamma}\cdot 
{\vec{\beta}}_{x}$), one arrives at the supermultiplet of spin range 2 .
These supermultiplets do not occur in the tensor product of two or more 
CPT self-conjugate doubleton supermultiplets, but they appear in the tensor 
product of two doubleton supermultiplets of spin 3/2.

The general lwvs for $j \geq 2$ \cite{gmz2} 
\eq
|\Omega\rangle = 
\xi^{[A_{1}}(1)\xi^{B_{1}]}(2)...\xi^{[A_{j}}(1) \xi^{B_{j}]}(2) 
|0\rangle\nn , \quad
|\Omega\rangle = 
\eta^{[A_{1}}(1)\eta^{B_{1}]}(2)...\eta^{[A_{j}}(1) \eta^{B_{j}]}(2) 
|0\rangle\nn
\en
lead to a supermultiplet with spin range 2 
and its CPT conjugate partner.
Obviously, the spin  content of these multiplets is independent of 
$j$. Only the AdS energies (conformal dimensions) get shifted, when 
$j$ is increased, which distinguishes these multiplets from their 
(super)symmetrized counterparts obtained from $|\Omega \rangle = 
\xi^{A_1}(1) \xi^{A_2}(1) ... \xi^{A_{2j}}(1) |0\rangle$, where the spins 
increase with $j$.

\section{Conclusions}

In this talk we have reviewed 
the unitary supermultiplets of the ${\cal{N}}=8$ $d=5$ 
anti-de Sitter 
superalgebra $SU(2,2|4)$ \cite{gmz1,gmz2} and  given a complete classification of the doubleton
supermultiplets of $SU(2,2|4)$. The doubleton supermultiplets do not have 
a smooth Poincar\'{e} limit in $d=5$. They correspond
to $d=4$ superconformal field theories  living on the boundary of 
$AdS_5$, 
where $SU(2,2|4)$ acts as the ${\cal{N}}=4$ extended superconformal algebra.
The unique CPT self-conjugate irreducible 
doubleton supermultiplet is simply 
the  ${\cal{N}}=4$ super Yang-Mills multiplet in $d=4$ \cite{mgnm}. However, there
are also chiral (i.e. non-CPT self-conjugate) doubleton supermultiplets 
with higher spins. The maximum spin range
of the general doubleton supermultiplet is 2. 
We have also reviewed the 
supermultiplets of $SU(2,2|4)$ that
can be obtained by 
tensoring two doubleton supermultiplets.  This class of supermultiplets 
has a maximal spin range of four and contains the multiplets that
are commonly referred to as ``massless'' in the $5d$ AdS sense including
the ``massless'' ${\cal{N}}=8$ graviton supermultiplet
in $AdS_5$ with   spin range two. Some of these supermultiplets 
were studied recently \cite{ferrara} using the language of
${\cal{N}}=4$  conformal superfields developed sometime ago
\cite{hst}.
We have pointed out that there exist some novel short supermultiplets of $SU(2,2|4)$ that have
spin range two and do not appear in the Kaluza-Klein spectrum of
IIB supergravity. These novel short supermultiplets do not occur in 
tensor products of the ${\cal{N}}=4$ Yang-Mills supermultiplet
with itself, but they can be obtained by tensoring higher spin chiral doubleton
supermultiplets. Both kinds of "massless" supermultiplets should be realized in the spectrum of
type IIB string theory on $AdS_5 \times S^5$ \cite{kallosh}.

{\bf Acknowledgements:}
I would like to thank  M. G\"{u}naydin and M. Zagermann for very enjoyable and fruitful collaboration and I. Bars, S. Ferrara, J. Maldacena and R. Kallosh for important  discussions.







\end{document}